\documentstyle[prl,aps,preprint,eqsecnum]{revtex}
\begin{document}
\draft

\title{ Wave Propagation in Gravitational Systems: Completeness of
Quasinormal Modes}

\author{E.S.C. Ching${}^{(1)}$, P.T. Leung${}^{(1)}$,  W.M.
Suen${}^{(2,3)}$ and K. Young${}^{(1)}$}
\address{${}^{(1)}$Department of Physics, The Chinese University of Hong
Kong, Shatin, Hong Kong}
\address{${}^{(2)}$McDonnell Center for the Space Sciences, Department
of Physics, Washington University,\\ St Louis, MO 63130}
\address{${}^{(3)}$National Center for Supercomputing Applications,
University of Illinois, Urbana, IL, 61801}

\date{\today}
\maketitle
\begin{abstract}
The dynamics of relativistic stars and black holes are often studied in terms
of the quasinormal modes (QNM's) of the Klein-Gordon (KG) equation with
different effective potentials $V(x)$.
In this paper we present a systematic study of the relation between
the structure of the QNM's of the KG equation and the form of $V(x)$.
In particular, we determine the requirements on $V(x)$ in order for
the QNM's to form complete sets, and discuss in what sense they form
complete sets.
Among other implications, this study opens up the possibility of using
QNM expansions to analyse the behavior of waves in relativistic
systems, even for systems whose QNM's do {\it not} form a complete
set.  For such systems, we show that a complete set of QNM's
can often be obtained by introducing an infinitesimal change in
the effective potential.

\end{abstract}

\setcounter{section}{0}
\section{Introduction}

Gravitational waves have been of theoretical interest since the
appearance of the general theory of relativity, and of experimental
interest for several decades.  The generation and propagation of
gravitational waves are often discussed in terms of a source region, a
strong-field and a weak-field near zone, and a local and distant wave
zone \cite{kip}.  For the purpose of this paper, a simpler point of
view will be useful for understanding: (i) a source region where the
radiation is generated, (ii) an intervening region with a non-trivial
geometry of the background spacetime, in which the propagation of
waves is significantly different from that in flat space, and (iii) an
``outside'' region where the spacetime can be taken to be flat, in
which the observer is located.  A close analogy is a molecule or
molecules [region (i)] radiating in an optical cavity [region (ii)].
The cavity significantly modifies the character of the electromagnetic
radiation observed outside [region (iii)].

The interest of this paper is on the effects of the intervening
nontrivial spacetime, i.e., the ``cavity'', in gravitational systems.
For example, consider radiation from a compact astrophysical
configuration, e.g., a binary neutron star system, embedded in a
galaxy.  The source region is the binary system, the ``cavity'' is the
galaxy, and the rest the ``outside''.  The presence of a ``cavity'' in
principle could have a significant influence on the radiation observed
from the outside.  The issue of a recent debate \cite{pp} is whether
the ``cavity", in this case, the galaxy, can suppress, by many orders
of magnitude, the emergence of radiation generated by the source
\cite{ppp}.  Another example is a particle radially plunging into a
black hole of mass $M$.  In this case the ``source region'' and the
``cavity" have the same size (as is true for a laser cavity).  The
``source'' is significant only within $\sim 10M $ of the plunge, and
the effective potential (in the perturbation theory of a Schwarzschild
hole \cite {bhqnm} ) representing the ``cavity'' is also significant
only within $\sim 10M$.  It is well-known that in this case the
``cavity'' modifies the observed radiation significantly.

Radiation from optical cavities is often discussed in terms of the ``modes''
of the cavity.  Because the waves escape to infinity, the
total energy in the finite part of the system decreases with time.  Thus,
these ``modes'' are characterized by complex frequencies $ \omega _{j}$, with
Im~$\omega _{j} <0 $.  The corresponding eigenfunctions are defined
by the outgoing wave condition at infinity.  These ``modes'' are in fact
quasinormal modes (QNM's).  The observation of the QNM's
from the outside gives directly information on the spatial structure
of the cavity (but not the structure of the source, unless a particular
model is assumed); e.g., in obvious notation,
$ \omega _{j} \sim j \pi c / L$, where $L$
is the length of a simple 1-dimensional optical cavity.

The same discussion extends to the gravitational case.
The QNM's of black holes \cite{bhqnm} and relativistic stars \cite{detweiler}
have been subjects of much study.  In numerical simulations, it is often
found \cite{numerwave} that the radiation observed in many black hole processes
is dominated by the QNM's of the background spacetime associated with the hole.
In view of the optical analogy, this
is not surprising --- the distant observer sees only the QNM's of the laser
cavity, but {\it not} the details of the source.
The fact that gravitational QNM's tend to be more ``leaky'' does not
change this qualitative understanding.

It is therefore particularly exciting that gravitational QNM radiation
may be observed by LIGO, VIRGO, and other detectors\cite{nature} in the next
decade. These observations will provide the
possibility of ``seeing'' directly interesting spacetime structures of various
gravitational systems, e.g.,  the spacetime around a black hole
--- in much the
same way as the spectrum of a laser permits a distant observer to infer the
geometry of the cavity.

Although the QNM's of black holes are quite well studied, the general
behavior of QNM's in more complicated gravitational systems has not
received much attention.  For example, in view of the possible
detection of gravitational waves, one would like to know: How much
would the QNM frequencies of a black hole be shifted by the
astrophysical environment that it is in, e.g., an accretion disk or
the host galaxy?  Can one calculate the shift perturbatively?  How
much would each QNM be excited in an astrophysical process, e.g., a
star plunging into a massive hole?  There are many such questions of
both theoretical and observational interest.

In an ongoing project, we study the general properties of QNM's
in gravitational systems.  In a recent
letter\cite{prl1}, we outlined the results obtained in studying
the QNM's of the Klein-Gordon (KG) equation
\begin{equation}
D \phi \equiv \left[ {\partial^2 \over \partial t^2}
      -{\partial^2 \over \partial x^2}
      +V(x)
\right] \phi (x,t) = 0  ,
\end{equation}
which is often used to describe
wave propagation in curved space,
with $V(x)$ describing the scattering of waves by the geometry.
In this paper, we provide the details of the results reported in\cite{prl1}.
We also present other related results and discuss the implications.

For a set of modes, the first question is whether they are
complete.  Completeness can mean several different things, and we shall
discuss the relations among the various senses of completeness in Sec.
III; for the moment we take the simple point of view that the QNM's are
complete if the evolution of the wavefunction can be expressed, in some
domain of time and space, as a sum
\begin{equation}
\sum_j a_j f_j(x) e^{-i \omega_j t}  ,
\end{equation}
where $f_{j}(x)$ are eigenfunctions with complex frequencies $\omega _{j}$. In
particular, if this holds at $t = 0,$ then an arbitrary function of $x$ within
a certain domain should be expressible as
\begin{equation}
\sum_j a_j f_j(x)  .
\end{equation}
Both of these properties can be discussed in terms of a Green's function
$G(x,y;t)$ for propagating a source at $y$ to an observation point at $x$ in
time {\it t.}
Our purpose is to elucidate under what conditions  would completeness hold.
This problem is nontrivial because usual proofs of completeness for the
normal modes of conservative systems rely crucially on
hermiticity. It is therefore particularly interesting that Price and
Husain \cite{price}
were able to give a model of relativistic stellar oscillations
in which the QNM's do form a complete set.
It was suggested\cite{price} that the completeness of the QNM's
stems either from the equality of the damping times of all the
QNM's, or from the absence of dispersion and backscatter in that model.
We shall see in this paper that these properties
are not the controlling factors of completeness.

There is a limit in which QNM's should behave trivially, like the
normal modes of hermitian systems, namely the limit in which the leakage goes
to zero; mathematically this means that the quality factor $Q_{j} \simeq$
Re $ \omega _{j}/(2\mid $Im $ \omega _{j}\mid)  \rightarrow \infty $.
In this limit, the QNM's become normal modes and are complete.
In the present paper we study the completeness of gravitational QNM's whose
quality factors are {\it not} large, and we
are interested in results that survive to all orders in the leakage,
which may be characterized by $1/Q_{j}$.
We show that the QNM's are complete in a broad class of models quite
independent of leakage, provided the
following three conditions hold:
\par
\begin{itemize}
\item[(i)] The effective potential $V(x)$ in (1.1) is everywhere finite,
and vanishes sufficiently rapidly,
in a sense to be defined in Sec. II, as $|x| \rightarrow \infty $;
\item[(ii)]  there are spatial discontinuities in $V(x)$  demarcating
the boundaries of a ``cavity"; and
\item[(iii)] consideration is limited to certain domains of space and time,
which will be spelt out below.
\end{itemize}

It is useful to comment heuristically on these conditions at the outset.
If (i) does not hold, i.e., if $V(x)$ has a significant tail at large $x$, then
it is possible for a disturbance originating at $y$ to propagate to a large
$x' \gg x$, and be scattered back to the observation point $x$. The time taken
would
be $O(x')$, and since $x'$ can be arbitrarily large, this could lead to a late
time tail in the wavefunction; crudely speaking this would be $t^{-\alpha}$ if
 asymptotically $V(x) \sim x^{-\alpha}$ (though this is modified in the
presence of a centrifugal barrier\cite{prl2}).  Such a
late time tail would not be described by a sum of discrete QNM's, which
must behave exponentially at large times.  Thus, the QNM's cannot be
complete for $V(x)$ having a tail.
The late time tail, and its relationship with the spatial asymptotics
of $V(x)$, is now thoroughly understood\cite{prl2}.

 A discrete sum over QNM's (like a Fourier series rather than a
Fourier integral) cannot possibly be complete over all space;
completeness, if it holds at all, can only be limited to the inside of
a finite interval.  The need for a natural demarcation of
this interval ``explains'' why discontinuities are needed. In the model
of \cite{price} there is such a discontinuity in the speed of wave
propagation at the model stellar surface.  This also explains why the
completeness relation (1.3) must be limited spatially.

The need to also limit the time domain in some cases can be understood from
the following example. Suppose the observation point is far from the source.
Then there will be transients propagating directly from the
source point $y$ to the observation point $x$ (i.e., the high
frequency components), without being much affected by
the potential; these have nothing to do with the QNM's.  Only past a
certain finite prompt time $t_{p} = O(x)$, when these transients have passed,
would the QNM's be complete.  This
case is most relevant for the discussion of gravitational waves seen by a
distant observer, and numerical experiments indeed show that quasinormal
ringing dominates only after the passage of transients.

It should be emphasized that the broad class of models satisfying
the above conditions and hence exhibiting completeness allow dispersion,
backscatter, as well as differences in the damping times of the QNM's,
showing that these are not essential ingredients.  This broad class of
models also allow the leakage to be not small; in other words, our results
are valid to all orders in $1/Q_{j}$.

There is one final technical hedge.  The QNM sums are in fact often divergent
series, but converge to the correct answer if regularized in a
standard way.  The simplest statement of
regulation is: keep only modes with Re $\omega_j >0$ by symmetry argument,
then replace all times $t$ by $t - i\tau$, with $\tau \rightarrow 0^{+}$.
We shall come to the need for this in Sec. II, and we shall also
demonstrate, through a numerical example in Sec. IV, that the
regulated series is useful in practice.

The general need for regularization is a major difference between the
completeness problem of the KG equation (1.1) and the wave equation
\begin{equation}
\left[  n^2 (z) {\partial^2 \over \partial t^2}
     -{\partial^2 \over \partial z^2} \right] \psi (z,t) = 0  .
\end{equation}
\noindent
where $n(z)$ may be regarded as a position-dependent refractive index in a
scalar model of electromagnetism. We have previously studied the completeness
of QNM's in the wave equation
case\cite{lly}.  In \cite{price}, the stellar oscillation model is
described by (1.4), with $n(z) =1 $ for $0 \le z \le L$ representing
the interior
of the star, and $ n(z)$ a non-zero constant for $ L < z < \infty $
representing space exterior to the star.  The boundary conditions are
$\psi (z=0) =0 $ and outgoing waves at infinity.

In this paper we work in terms of the KG equation
(1.1) since it is more realistic as a model for wave propagation
in curved space, and has been much studied for this purpose.
For example, in a static,
 spherically symmetric spacetime,
\begin{equation}
ds^{2} = g_{tt}(r)dt^{2} + g_{rr}(r)dr^{2}
+ r^{2}(d\theta ^{2} + \sin ^{2}\theta  d \varphi ^{2})  ,
\end{equation}
a KG scalar field $\Phi $ can be expressed as
\begin{equation}
\Phi  =\sum_{lm} {1\over r} \phi _{\l m}(x,t) Y_{lm}(\theta ,\varphi ) .
\end{equation}
The evolution of each $\phi _{\l m}(x,t)$ is given by (1.1) with the
effective potential
\begin{equation}
V(x) = - g_{tt} {l(l + 1)\over r^{2}} -{1\over 2r} {g_{tt}\over g_{rr}}
({1\over g_{tt}} {\partial  g_{tt}\over \partial r}-
{1\over g_{rr}} {\partial  g_{rr}\over \partial r}).
\end{equation}
The variable $x$ in the KG equation is related to the circumferential
radius $r$ by
\begin{equation}
x = \int \sqrt{-{g_{rr}\over g_{tt}}} dr .
\end{equation}
The particular case of (1.1) arising from a Schwarzschild spacetime is
well studied.  The Maxwell
field and the linearized gravitational waves satisfy the same equation with
slightly different potentials $V(x)$ \cite {bhqnm}.

The wave equation (1.4) is related to the KG equation (1.1)
by a transformation:
\begin{eqnarray}
\nonumber
     dx/dz && = n(z)   ,\\
     \psi  && = n^{-1/2} \phi    ,
\end{eqnarray}
with the potential related to $n(z) $ by
\begin{equation}
    V = (2n^3)^{-1}(d^2 n/dz^2) -(3/4n^4)(dn/dz)^2 \, .
\end{equation}
Hence the results obtained in one case can be restated in the other,
up to possible complication coming from regularization, which is
investigated in this paper.  As we shall see later, discontinuity is a
crucial feature in the consideration.  We note that $V(x)$ with a
discontinuity in its $p$-th derivative maps into $n(z)$ with a
discontinuity in its $(p+2)$-th derivative.

Our results for discontinuous potentials, in addition to clarifying the
conditions for completeness, are significant in two ways.
First, they are directly applicable to physical models with a
discontinuity, e.g., a
stellar surface as in the model of \cite{price}.  Second,
any smooth potential can be
approximated, to arbitrary precision, by a discontinuous one.  In fact, any
numerical scheme that employs  finite spatial differencing has in effect
replaced $V(x)$ by one with a discontinuity in a high order derivative.  Thus
the complete set of QNM's of the discontinuous system can be used as an
effective
calculational tool for the original system. Further remarks on this aspect
are given in Sec. V, while detailed studies along this line
will be reported elsewhere.

In Sec. II we develop the formalism via the Green's function in
the complex frequency plane, and demonstrate that for a class
of discontinuous potentials $V$,
the Green's function is given by a sum of QNM's.
The conceptual analogy with optical cavities extends to a
close parallel in the mathematical treatment with the optical
case \cite{lly}, but there
are technical differences which are stated in this paper.
Sec. III spells out more clearly the different senses of completeness and also
discusses briefly the extension to KG equation with a non-zero mass.
In Sec. IV, we verify numerically the completeness relation
in a model problem.  The paper ends with some discussions and conclusion.

\newpage

\section{ The Green's function}
\subsection { Green's function in frequency plane}
We model the propagation of waves in curved space by the KG equation (1.1).
\noindent The potential $V(x)$ is assumed to be bounded and positive, and
represents the scattering of waves by the background. Generically, we take
$V_{o}\equiv  \lim_{x\rightarrow \infty } V(x)=0$. The case $V_{o} \neq 0$,
which represents a massive field, will be discussed briefly in Sec. III.
The spatial coordinate $x$ often
 represents a radial variable [cf. (1.8)], so we shall first  consider  a
half-line problem $(x \ge  0 )$ with the regular boundary condition
$\phi (x=0,t) = 0$.
The full line problem ($- \infty < x < \infty $), e.g.,
$x$ being the tortoise coordinate in the Schwarzschild case, will be studied
later.  The
QNM's are eigen-solutions $\phi (x,t) = \tilde{\phi} (x) e^{-i\omega t }$  of
(1.1), where $ \tilde{\phi}$ satisfies
\par
\begin{equation}
\tilde{D} \tilde{\phi} \equiv
[- \omega ^{2} -{\partial ^{2}\over \partial x^{2}} + V(x)] \tilde{\phi} (x) =
0,
\end{equation}
\noindent with $ \tilde{\phi}(x=0)=0 $ and the outgoing wave boundary condition
at infinity. The eigenfunctions and eigenvalues are defined as $
\tilde{\phi}(x)
=f_j(x)$ and $\omega=\omega_j$.
\par
The Green's function $G(x,y;t)$ for the system is defined by
$D G = \delta (x - y) \delta (t)$ with the initial condition
$G = 0$ for $ t  \le  0$ and the boundary conditions that
(i) $G = 0$ for either $x = 0$ or $y = 0$,
and (ii) the outgoing wave condition as either $x \rightarrow \infty $ or
$y \rightarrow \infty $.
The corresponding Green's function in the frequency domain is
\begin{equation}
\tilde{D} \tilde{G}(x,y;\omega ) = \delta (x -  y).
\end{equation}
The strategy is to express $G$ in terms of $\tilde{G}$, and attempt to
close the
contour by a large semicircle in the lower half $\omega$ plane.

Introduce two auxiliary functions
$f(\omega ,x)$ and $g(\omega ,x)$, which  are  the
solutions to the homogeneous time-independent KG equation
$\tilde{D} f(\omega ,x) =\tilde{D} g(\omega ,x) = 0$
with the boundary conditions
$f(\omega ,x=0) =  0$; $f'(\omega ,x=0) = 1$ \cite{origin} and
$\lim_{x\rightarrow \infty } \left [g(\omega ,x)  \exp (-i \omega  x) \right]
= 1$. With these auxiliary functions the Green's function is then given by
\par
\begin{eqnarray}
\nonumber
\tilde{G}(x,y;\omega )&=& f(\omega ,x) g(\omega ,y) / W(\omega )
\ \ \ \ {\rm  for }\ \ 0 < x < y,\\
&=& f(\omega ,y) g(\omega ,x) / W(\omega ) \ \ \ \ {\rm for }\ \ 0 < y < x,
\end{eqnarray}
\noindent where the Wronskian
$W(\omega ) = g(\omega ,x) f'(\omega ,x) - f(\omega ,x) g'(\omega ,x)$
($'=d/dx$) is independent of $ x$.
$\tilde{G}(x,y;\omega )$, of course, may be singular in  $\omega$
at the singularities of $f$  and $g$; otherwise,
it  is analytic except at zeros of $W(\omega )$.
 At these zeros $f$
and $g$ are proportional to each other, so each of them
satisfies the regular boundary condition at $x = 0$, {\it and}  the  outgoing
wave
boundary condition as $x \rightarrow \infty $. Hence these frequencies
are exactly the QNM frequencies $\omega_j$ and $f(\omega _{j},x)=f_j(x)$.
We shall further assume for simplicity that these zeros are simple
 \cite{pole}, so that the residues
of $\tilde{G}(x,y;\omega)$ at these zeros are given by
\begin{equation}
K_{j} = f(\omega _{j},x) g(\omega _{j},y) / [\partial W(\omega =\omega
_{j})/\partial \omega ].
\end{equation}
\noindent
 Multiple zeros can be handled readily. To study the physical meaning of the
denominator $\partial W/\partial \omega$,
 start with the defining equation for  $f(\omega_{j},x)$
and $g(\omega,x)$. The usual manipulations
lead to
\begin{equation}
(\omega^2-\omega_{j}^2) \int_0^X dx \ f(\omega_j,x)g(\omega,x)=
g(\omega,x)f'(\omega_j,x)-g'(\omega,x)f(\omega_j,x)|_0^X,
\end{equation}
where the integral is taken along any contour from $x=0$ to $x=X$. Since both
$f(\omega_j,x)$ and $g(\omega,x)$ are outgoing waves at $x=X$, the right-hand
side of
(2.5) becomes
\begin{equation}
i(\omega_j-\omega)f(\omega_j,X)g(\omega,X)-
\left[g(\omega,0)f'(\omega_j,0)-g'(\omega,0)f(\omega_j,0)\right],
\end{equation}
as $X \rightarrow \infty$. Differentiating (2.5) with respect to $\omega$ and
taking $\omega \rightarrow \omega_j$ then gives
\begin{equation}
\lim_{X \rightarrow \infty} \int_0^X dx \ f(\omega_j,x)g(\omega_j,x)+
{i \over 2\omega_j }f(\omega_j,X)g(\omega_j,X)
=-{1 \over 2\omega_j}{\partial W(\omega=\omega_j) \over \partial \omega}.
\end{equation}

\noindent Since $f$ and $g$ are proportional to each other
at  these  poles,  the  residues  can  be
expressed in terms of a generalized norm of the QNM's
\begin{equation}
\ll f_{j}\mid f_{j} \gg \ \equiv \lim_{X \rightarrow \infty}
 \int_0^X dx \ f(\omega_j,x)f(\omega_j,x)
+{i \over 2\omega_j }f(\omega_j,X)f(\omega_j,X)
\end{equation}
\noindent as follows
\begin{equation}
K_{j} = - {f_{j}(x) f_{j}(y)\over 2\omega _{j}\ll f_{j}\mid f_{j}\gg }.
\end{equation}
The norm (2.8) has been introduced in other contexts \cite{lly,zel,lam} and
has the following significant properties:
\begin{itemize}
\item[(i)] it invovles $f^2$ rather than $|f|^2$, and is therefore in general
complex;
\item[(ii)] there is a surface term;
\item[(iii)] although each of the two terms on the right of (2.8)
does not have a limit, the limit exists for the sum; and
\item[(iv)] it reduces to the usual norm if the leakage goes to zero.
\end{itemize}

The property (iii) is easily verified by differentiating the right hand side
of (2.8) with respect to $X$, and using the outgoing wave condition,
which also applies to $f$  at $\omega=\omega_j$.
The appearance of the surface term is closely connected
with the fact that the momentum  operator is not hermitian.
While (2.8), with the integral taken along the real axis,
is often convenient for actual evaluation,  we may choose a deformed
contour $L$ running from $x=0$ to $x=-\infty$. The first part $L_1 =(0,a)$,
and the second part $L_2$ is shown in Fig. 1. On $L_2$, $f(\omega_j,x)$ is
defined as the outside solution (which is analytic in $x$ for $x>a$),
analytically continued to the complex plane. Along this
contour, $f(\omega_j,x) \rightarrow 0$ as $x \rightarrow -\infty$, and we get,
more compactly
\begin{equation}
\ll f_{j}\mid f_{j} \gg \ \equiv  \int_{L} dx f(\omega _{j},x)^{2} dx,
\end{equation}
 making the formal analogy to the hermitian case more transparent.
However, analytic continuation out into the complex $x$ plane is not strictly
necessary if the surface term is retained, as in (2.8).

Next write $G$ in terms of $\tilde{G}$; then upon closure of the contour
in the lower half $\omega$ plane, one sees that
\begin{equation}
G(x,y;t) = {i\over 2} \sum^{}_{j} {f_{j}(x) f_{j}(y) e^{- i\omega _{j}t}
\over \omega _{j}\ll f_{j} \mid f_{j}\gg } + I_{c} + I_{s}.
\end{equation}
In (2.11), the sum comes
from the zeros of the Wronskian $W(\omega)$ inside the semicircle, $I_{c}$
is the  integral along a semicircle at
infinity, and $I_{s}$ comes from the singularities of $f(\omega,x)$ and
 $g(\omega,x)$ (see Fig. 2).
 The crux of the proof of completeness then lies in
(i) determining the conditions on $V(x)$
under which $f$ and $g$ have no singularities in the $\omega$ plane,
in which case $I_{s}=0$; and (ii)
 showing (in the next subsection) that $I_{c}$ vanishes if there is a
discontinuity in the potential $V(x)$ at some $x = a >0.$
Under these conditions, the Green's function can be represented exactly in
terms of the QNM's,
which then establishes these QNM's as a sufficient basis for discussing the
dynamics. The behavior is a discrete sum of exponentials, and in particular,
there
would be no power-law dependence at large times.

Since QNM's appear in pairs at $\omega=\omega_j$ and $\omega=-\omega_j^*$, it
is
convenient to rewrite the QNM sum (when $I_s=I_c=0$) as
\begin{equation}
G(x,y;t)={\rm Re} \left[i \sum_{j>0}{f_j(x)f_j(y) e^{-i \omega_{j} t}
 \over \omega \ll f_j|f_j \gg}\right],
\end{equation}
where the notation $j>0$ is a shorthand for Re $\omega_j >0$. This latter form
is
somewhat more convenient for regularization when the sum diverges.
The corresponding statements for $\dot{G}(x,y;t)$, and consequently a QNM
representation of $\delta(x-y) = \dot{G}(x,y;t=0^+)$, follow relatively
simply and will be discussed in Sec. III.

Now to determine the conditions under
which $f(\omega,x)$ and $g(\omega,x)$ have no singularities in $\omega$,
we appeal to well known results in the quantum theory of scattering.
The defining equation for $ f$ and $g$ is identical to the Schr\"{o}dinger
equation
for a particle with mass =  1/2 and energy $= \omega ^{2}$ moving
in a potential well $V(x)$.  It  has been
proved \cite{newton} that $f$ and $g$ are analytic functions of $\omega $ if
the potential is bounded and ``has no tail", in the sense that
\begin{mathletters}
\begin{eqnarray}
&&\int^{\infty }_{0} dx \ x \mid V(x)\mid  < \infty , \\
&&\int^{\infty }_{0} dx \ x \ e^{\alpha x}\mid V(x)\mid  < \infty \ \
  {\rm  for \ any}\  \alpha  > 0  .
\end{eqnarray}
\end{mathletters}Note that if condition (2.13b) is violated for
some $\alpha  > \alpha _{o} >0$, then
$g(\omega ,x)$ may not be analytic for Im $\omega  < -\alpha _{o}$.
The singularities that appear for potentials that have a tail, e.g.,
inverse-power-law potentials,
have been studied, with a focus on understanding the associated
late time behavior\cite{prl2}.

It then remains to investigate the behavior of $\tilde{G}$ on
the large semicircle.

\subsection { Asymptotic behavior of Green's function}
We first give a simple (and not strictly rigorous) derivation to highlight the
essential ideas.
The asymptotic  behavior  of $\tilde{G}$  at  high
frequencies can be determined using the WKB method.
Let $\tilde{\phi} (x) \equiv  \exp [i S(x)]$  be  a
solution of the time-independent KG equation; then  in lowest  order
approximation
\begin{equation}
S(x) = \pm \int^x k(x') dx'  ,
\end{equation}
where the position-dependent wave number $k(x)$ is given by
$ k(x) = [ \omega ^{2} - V(x) ]^{1/2}$.
The two auxiliary functions $f$ and $g$,
and hence the Green's function $\tilde{G}$,
can be obtained in  terms of  $\tilde{\phi} (x)$ and $k(x)$.

Consider a potential with a discontinuity at $x=a$.
In this situation the WKB approximation breaks down at the discontinuity;
 however, one can join the two  approximate  solutions  across the
 discontinuity using  the  reflection coefficient
\par
\begin{equation}
R = {S'(a^{-}) - S'(a^{+})\over [S'(a^{-})]^* + S'(a^{+})} \ .
\end{equation}
The discontinuity can be in any derivative of $V$, and $R$ behaves as some
inverse power of $\omega$ as $|\omega| \rightarrow \infty$. (This will be spelt
out precisely below.) We divide the discussion into two cases:
(a) $ 0 < y < a < x$;
(b) $ 0 < y \le  x < a$.
Referring to the discontinuity at $x = a$ as the stellar surface for
simplicity,
we may say that case (a) is relevant when gravitational waves from a
source inside the star is detected by a distant observer outside the star,
while case (b) is relevant for discussing the internal dynamics of
the star.
\par
\medskip
For case (a), it is straightforward to show that
\par
\begin{equation}
\tilde{G}(x,y;\omega ) \simeq  \frac{(1+R)
\sin [I(0,y)]e^{iI(a,x)}}{ \sqrt{k(x)k(y)} [e^{-iI(0,a)} + Re^{iI(0,a)}]}\ ,
\ \ \ y < a < x,
\end{equation}
\noindent where
\begin{equation}
I(u,v) =\int ^{v}_{u} k(x) dx \approx  \omega (v- u).
\end{equation}
Now consider $\tilde{G}$ on the semicircle $\omega  = \omega _{R} + i\omega
_{I} = C e^{i\theta }, \pi  < \theta  <
2\pi $, with $C \rightarrow  \infty $.  As $\omega _{I} \rightarrow  - \infty
$, both the numerator and the denominator of $\tilde{G}$ are
dominated by the term proportional to $R$, and
\begin{equation}
\tilde{G}(x,y;\omega ) e^{- i\omega t} \simeq
{(1+R)e^{-i \omega(t-x-y+2a) } \over R \omega}.
\end{equation}
\noindent As $C \rightarrow  \infty $, this vanishes for
 $t > t_{p} \equiv \max(x + y-2a,0)$, since  $R$  varies as an
inverse power of $\omega$.
Thus we have proved (modulo a technical complication to be mended below), that
for a discontinuous potential, completeness in the sense (1.2) holds for
 $t > t_{p}(x,y)$.

Next consider case (b).  Instead of (2.16), one now has
\begin{equation}
\tilde{G}(x,y;\omega ) \simeq  \frac{\sin [I(0,y)][e^{-iI(x,a)} +
Re^{iI(x,a)}]}{ \sqrt{k(x)k(y)} [e^{-iI(0,a)} + Re^{iI(0,a)}]}\ .
\end{equation}

\noindent Again, both the numerator and the denominator in $\tilde{G}$ are
dominated by the term
proportional to $R$, and
\begin{equation}
\tilde{G}(x,y;\omega ) e^{- i\omega t} \simeq {e^{-i \omega(t+x-y) } \over
\omega},
\ \ \ \ y \le x < a.
\end{equation}
As $C \rightarrow  \infty $, this vanishes for all $t > 0.$  Thus completeness
is again proved.
\par
\medskip
The above derivation shows the essence of the proof, which
relies on the vanishing of an exponential factor
 $e^{- \mid \omega _{I}\mid \sigma }$ where $\sigma  > 0$ (say $\sigma  =
t - x - y+2a)$, when $\omega  = C e^{i\theta}, \pi  < \theta  < 2\pi $ and $C
\rightarrow  \infty $.  This derivation however
must be mended in the domain $\mid \omega _{I}\mid  = O(\log C)$, i.e., within
an angle $\Delta \theta \sim \log C/C$ of the
real axis, in which  $e^{- \mid \omega _{I}\mid \sigma }$ behaves only as a
power $C^\sigma$, rather than as an exponential\cite{jordan}.  The problem is
closely related to the QNM's, which lie precisely
in this domain.
\par
\medskip
To discuss this in a somewhat more general context, consider a potential
with a discontinuity in $d^{p}V(x)/dx^{p}$ at $x = a$, for some
$p \ge  0.$  Then as $\omega \rightarrow  \infty $,
$R \sim A \omega ^{- q}$, where $q = p + 2$.  We first determine the
asymptotic position of the QNM's, which are given by the zeros of the
denominator  $\tilde{G}$ in (2.16):
\par
\begin{equation}
e^{-iI(0,a)} + R e^{iI(0,a)} \simeq  e^{-i\omega a} + A \omega^{-q} e^{i\omega
a} = 0,
\end{equation}
\noindent with the solution
\par
\begin{equation}
\omega _{j}a \simeq j \pi -i\left[q \log(j \pi)- \log(-a^qA) \right]/2,
\end{equation}
\noindent where $j$ takes an integer values.  The distribution of QNM's is
shown
schematically in Fig. 3.

The derivation sketched above  relies on the fact that the integrand vanishes
on the large semicircle as $C \rightarrow \infty$. However, for $\omega
\simeq \omega_j$, one can readily obtain the following estimate for
 the integrand:
\begin{equation}
|\tilde{G}(x,y;\omega ) e^{- i\omega t}|
 \sim {|R|^{(t-x-y)/2a} \over |\omega|}
 \sim {|\omega^{-q(t-x-y)/2a}| \over |\omega|},
\end{equation}
for both case (a) and case (b). In order to justify the derivation for
the completeness of QNM, one could restrict $t > x+y$,
but this would greatly limit the usefulness of the completeness relation.
However, it will be shown in the next subsection that this difficulty can be
surmounted by using a standard regularization scheme.

\par
\medskip
The situation for $t<x+y$ in case (b), which does not give convergence, is
markedly
different from the case of the wave equation with a step discontinuity,
which was considered by us in the context of optics \cite{lly},
and also by Price and Hunain as a model for gravitational waves \cite{price}.
 In that case, the analog of the potential $V(x)$ is
$\omega ^{2}\epsilon (x)$, where $\epsilon(x) \equiv n(x)^2 $ is the dielectric
constant.  Any discontinuity in $\epsilon (x)$ is
therefore amplified as $\omega  \rightarrow  \infty $ by two powers;
specifically, if there is a
discontinuity in $d^{p}\epsilon (x)/dx^{p}$, then $R \sim A \omega ^{- q}$
at high frequencies, with $q = p$. The difference by two powers also follows
from the mapping (1.10). Thus for $p = 0$, $\tilde{G}$ is
sufficiently bounded for  the proof of completeness to go through \cite{lly},
and the statements about the possibility of expressing
  $G$ and $\partial G/\partial t$
as sums over QNM's are strictly valid for the wave equation\cite{convergent}.
If $p \ge  1,$ then the same problem arises even for the wave equation.
\par
\medskip
However, both the case of the KG equation, and the case of the wave
equation with $p \ge  1,$ can be rectified by a simple
scheme of regularization.

\subsection{ Regularization}
First, it follows directly from the reflection symmetry relation
$\tilde{G}(x,y;-\omega^*)=\tilde{G}(x,y;\omega)^*$ that

\begin{equation}
G(x,y;t) = 2\,{\rm Re}\int_{0}^{\infty}
{d \omega \over 2 \pi}e^{-i \omega t} \tilde{G}(x,y; \omega).
\end{equation}
 As the integrand is well behaved at both endpoints, we can multiply it by an
extra factor $e^{- \omega \tau}$, $\tau > 0$, and then take the limit
$\tau \rightarrow 0^{+}$ afterwards, i.e.,

\begin{equation}
G(x,y;t) = 2 \,{\rm Re} \lim_{\tau \rightarrow 0^+} \int_{0}^{\infty}
{d \omega \over 2 \pi}e^{-i \omega (t-i\tau)} \tilde{G}(x,y; \omega).
\end{equation}
This is equivalent to assigning $t$ an infinitesimal negative imaginary
part, and provides sufficient regulation on the part of
the contour with $|\omega_I|$ small to make the integral vanish as the contour
expands to infinity in the lower half plane.

Second, we deform the integration contour in (2.25) into a
sequence of expanding rectangular contours
 $\Gamma _{n}=\Gamma_{n1} \bigcup \Gamma_{n2} \bigcup
\Gamma_{n3}$ in the lower half $\omega$ plane as shown in Fig. 4.
However, instead of letting the width of the contour $\Gamma_{n}$ expand
continuously to infinity (which would hit the QNM's), each $\Gamma_{n}$ is
defined to cut roughly mid-way between $\omega _{n}$ and $\omega _{n+1}$,
and eventually we take the limit $n \rightarrow \infty $.
The height $\Omega_n$ of the contour is chosen to be proportional to $n$.
As a consequence, $G$ can be expressed as a sum over QNM's lying within
the rectangle formed by the real axis and $\Gamma_{n}$\cite{0mode},
plus an integral along the contour $\Gamma _{n}$:
\begin{equation}
G(x,y;t) =  2\,{\rm Re}  \int_{\Gamma_{n}}
      {d \omega \over 2 \pi}e^{-i \omega (t-i\tau)} \tilde{G}(x,y; \omega)
      +{\rm Re} \lim_{\tau \rightarrow 0^+}
\left[i \sum_{j>0}^{n}{f_j(x)f_j(y) e^{-i \omega_{j} (t-i\tau)} \over
\omega_j \ll f_j|f_j \gg}\right].
\end{equation}

Third, provided $V(x)$ does not have a tail, there are no cuts,
and  $\tilde{G}(x,y; \omega)$ is real along the imaginary axis,
which also follows from the reflection symmetry of $\tilde{G}$.  Thus
the integral along $\Gamma_{n1}$ is purely imaginary and does
not contribute to $G$.

For the integral along $\Gamma_{n2}$, the original argument sketched in the
previous subsection
is valid: when $\Omega_n$ is sufficiently large, the integrand goes as
$\exp (-\mid \omega _{I}\mid \sigma )$, $\sigma  > 0,$ and therefore vanishes
in the limit $\Omega_n \rightarrow \infty$; the condition $t > t_{p}$ for case
(a) is
necessary in order that $\sigma  > 0.$

Next, divide each contour $\Gamma_{n3}$ into two parts, with
$\mid \omega _{I}\mid $ large in one part, and $\mid \omega _{I}\mid $ small
in the other. In the first part, the integral also vanishes
as the contours expand to infinity, for the same reason mentioned previously.
The complication, as indicated earlier, lies with the second part, where
$\mid \omega _{I}\mid $ is not large, and therefore does not provide an
exponential damping factor. However, in this part the integrand is at most a
power of $\omega$ and the regulating factor $e^{- \omega \tau}$ causes it to
vanish as the contours expand to infinity.

As the contribution from the contour $\Gamma_n$ vanishes when $n \rightarrow
\infty$, the Green's function $G(x,y;t)$ can be decomposed into a regulated
sum of QNM's, namely
\begin{equation}
G(x,y;t)={\rm Re} \lim_{\tau \rightarrow 0^+}
\left[i \sum_{j>0}{f_j(x)f_j(y) e^{-i \omega_{j} (t-i\tau)} \over
\omega_j \ll f_j|f_j \gg}\right].
\end{equation}
Thus the prescription is, very simply, to (i) consider only $\omega _{R} > 0$
and
(ii) where necessary give $t$ a small negative imaginary part $\tau$.
 (The exceptional case where some QNM's
lie exactly on the imaginary axis is readily handled\cite{0mode}.)

The physical meaning of the regularization is as follows. The Green's function
$G(x,y;t)$ is a well defined object, independent of the size of the
contour $\Gamma $.  We attempt to write it as the sum of a prompt part arising
from the integral on a large semicircle (or other large contour linking
$\omega  = -i\infty$ to $\omega  = +\infty$), and the sum over QNM's.
When the QNM frequencies extend to
infinity, and there are insufficient powers of $1/\omega $ in the integrand,
each of the two contributions individually diverges as the contour expands to
infinity\cite{jordan}.  Physically, in these circumstances, high frequency
QNM's cannot be
cleanly separated from the prompt contribution, which is hardly surprising.
However, the device of letting $t \rightarrow  t- i\tau$ effects such a clean
separation which
is practically useful, as illustrated in the numerical examples in Sec. IV.

Mathematically, the regulator $e^{- \omega _{j}\tau}$ can be replaced by
any other regulator $I_{j}(\tau)$ satisfying
 $I_j(\tau) \rightarrow 1$ as $\tau \rightarrow 0$
and  $I_j(\tau) \rightarrow 0$ as $j \rightarrow \infty$.
One possible alternative is
\par
\begin{eqnarray}
\nonumber
I_j(\tau) &=& (N-j+1)/N, \ \ \ j \le N,  \\
	  &=& 0,    \ \ \ j > N,
\end{eqnarray}
where $N$ is the least integer greater than $1/\tau$; this corresponds to
defining the divergent series by a Cesaro sum
\cite{hardy}.

In practice, one would use a small value of $\tau$ rather than take $\tau
\rightarrow  0^{+}$.
Consider, for example, the regulator $e^{- \omega _{j}\tau}$ and using
$G$ to propagate given initial data.  Leaving $\tau$ finite incurs an error
only
for those modes with $|\omega _{j}| {\buildrel > \over \sim} 1/\tau$,
and hence only affects the resultant
wavefunction on length scales below $\Delta \sim  1/|\omega _{j}|
{\buildrel < \over \sim} \tau$.  Thus, so long as only a
finite spatial resolution is required, not taking $\tau \rightarrow  0^{+}$ has
little effect,
as demonstrated in the numerical example.

\subsection{Full line problem}

The proof of completeness can be generalized to include cases with multiple
 discontinuities and also to problems on a full line:
 $-\infty < x < \infty.$  The latter is physically relevant if the background
spacetime is due to a Schwarzschild
black hole of mass $M$. Under the transformation (1.8), or more explicitly
\begin{equation}
x=r+2M \ln({r \over 2M}-1),
\end{equation}
 the event horizon ($r=2M$)  maps into $x \rightarrow -\infty$, while $r
 \rightarrow +\infty$ maps to  $x \rightarrow +\infty$.
The main idea is still to construct the Green's function $\tilde{G}$ from the
two
 homogeneous solutions   of the KG equation and then examine its
high frequency behavior using the WKB approximation. Consider a
full line problem
 where there are $n$ discontinuities in $V(x)$ at $x = a_1,a_2,...,a_n$.
We show only the case $a_1 \le y \le x \le a_n$; the case where the observation
point is outside $(a_1,a_n)$ can be demonstrated similarly.
The QNM's are defined by (2.1) with the outgoing waves conditions
at $x\rightarrow\pm\infty$. For simplicity we shall assume that
$\lim_{x\rightarrow\pm \infty} V(x)  = 0$,
and consequently
\begin{equation}
\lim_{x\rightarrow\pm\infty} f_j(x)  \sim \exp(i \omega \mid x \mid).
\end{equation}
In addition, we assume that $V(x)$ has no tail on either side,
in the sense of (2.13).

Introduce two auxiliary functions $ g_-(\omega,x)$ and $g_+(\omega,x)$ which
are the solutions
to the time-independent KG equation $\tilde{D}g_{\pm}=0$,
 with the boundary conditions
$\lim_{x\rightarrow\pm\infty}\left[ g_\pm (x) e^{\mp i\omega x}\right]=1$.
As before, the Green's function in frequency space is
\begin{eqnarray}
\nonumber
\tilde{G}(x,y;\omega) &=& g_+(\omega,x) g_-(\omega,y)/W(\omega) \; \; \; {\rm
for}\; y<x,\\
	& = &g_+(\omega,y) g_-(\omega,x)/W(\omega) \; \; \; {\rm for}\; x<y,
\end{eqnarray}
where the Wronksian
$W(\omega) = g_+(\omega,x) g_-' (\omega,x) -g_+' (\omega,x) g_-(\omega,x)$
is again independent of $x$.
Since $V(x)$ has no tail, $g_\pm(\omega,x)$ are analytic
in $\omega$, and hence $\tilde{G}$ is also analytic except at the zeros of
$W(\omega)$.
If one can show that the integral along the semicircle at infinity vanishes,
the
QNM's will form a complete set, as described in Sec. IIB.

Let us examine the asymptotic behavior of $g_\pm (\omega,x)$
along a large semicircle in the half plane using the WKB approximation.
Assume that $a_1 < a_2 < ...< a_n$, then for
$x > a_n$
\begin{equation}
g_+(\omega,x)  \simeq {\rm exp}[i I(a_n,x)]g_+(\omega,x=a_n).
\end{equation}
Now $g_+(\omega,x)$ in general will consist of two counter-propagating
waves for $a_j \le x \le a_{j+1} $ and can
be expressed as follows:
\begin{equation}
g_+(\omega,x)  \simeq A_j {\rm exp}[i I(a_j,x)] + B_j {\rm exp}[-i I(a_j,x)].
\end{equation}
The coefficients $A_j$ and $ B_j$ can be obtained recursively from the relation
\begin{eqnarray}
\left( \begin{array}{c} A_{j} \\ B_{j} \end{array} \right) & =&
   \left( \begin{array}{cc} M_{11}(j) & M_{12}(j) \\ M_{21}(j) & M_{22}(j)
\end{array} \right) \times \\
& &\left( \begin{array}{cc} \exp[i I(a_{j-1},a_j)] & 0 \\ 0 & \exp[-i
I(a_{j-1},a_j)]  \end{array} \right)
\left( \begin{array}{c} A_{j-1} \\ B_{j-1} \end{array} \right) \nonumber
 \end{eqnarray}
The boundary condition is $A_n =g_+(\omega,x=a_n)$ and $B_n = 0$.
The transfer matrix ${\bf M}(j)$ joining the wavefunction across the
discontinuity at $x = a_{j} $
can be obtained by matching the boundary conditions at  the specified point.
At high frequencies
\begin{equation}
M_{11}(j) = M_{22}(j) \simeq 1,
\end{equation}
and
\begin{equation}
M_{12}(j) = M_{21}(j) \simeq -R_j  ,
\end{equation}
 where $R_j$ is the reflection coefficient
defined by (2.15) and evaluated at  $x = a_{j}$.

The asymptotic form of $g_-(\omega, x)$ can be similarly expressed in terms of
two counter-propagating waves for $a_j \le x \le a_{j+1} $,
\begin{equation}
g_-(\omega,x)  \simeq C_j {\rm exp}[i I(a_j,x)] + D_j {\rm exp}[-i I(a_j,x)],
\end{equation}
where the coefficients $C_j$ and $D_j$ obey the same recursion relation as
$A_j$ and $B_j$, but with  the initial conditions
\begin{equation}
\left( \begin{array}{c} C_{1} \\ D_{1} \end{array} \right)  =
   \left( \begin{array}{cc} M_{11}(1) & M_{12}(1) \\ M_{21}(1) & M_{22}(1)
\end{array} \right)
  \left( \begin{array}{c} 0 \\ g_-(\omega,x=a_1) \end{array} \right)
 \end{equation}
Then for $x \le a_1$
\begin{equation}
g_-(\omega,x)  \simeq {\rm exp}[-i I(a_1,x)]g_-(\omega,x=a_1).
\end{equation}

 We now demonstrate the completeness of QNM's for $x \in (a_1, a_n)$ as
follows.
As $\omega \rightarrow \infty$ in the lower half plane, the two auxiliary
functions
can be simplified by keeping only waves which grow exponentially in their
expansions, and consequently
\begin{equation}
g_+(\omega,x)  \simeq -R_n{\rm exp}[-i I(a_n,x)]g_+(\omega,x=a_n),
\end{equation}
\begin{equation}
g_-(\omega,x)  \simeq -R_1{\rm exp}[+i I(a_1,x)]g_-(\omega,x=a_1).
\end{equation}
The Green's function is thus given by
\begin{equation}
{\tilde G}(x,y;\omega) \simeq {\rm exp}\{ i[I(a_1,y)+I(x,a_n)-I(a_1,a_n)]\}
/(2i \omega)
\end{equation}
for $a_1 < y \le x < a_n$, and vanishes as $\omega_I \rightarrow -\infty$. The
completeness of QNM's then follows.
There is the same problem in the domain $|\omega_I| = O(\log |\omega_R|)$,
and the same regularization needs to be applied.

\newpage

\section{ Different Senses of Completeness}
 We are now in a position to specify precisely what we mean by
completeness, and to summarize our results in terms of these definitions.
Completeness can be expressed in different forms and the meanings may or may
not
be the same.  The first sense $(C1)$ shall mean the
validity of the following expression, in a certain domain of $x,$ $y,$ $t$, for
the Green's function $G$ for problems with an outgoing boundary condition:
\par
\begin{equation}
C1: \ \ \ G(x,y;t) = \lim_{ \tau \rightarrow 0} {\rm Re}
\left[i \sum_{j>0}{f_j(x)f_j(y) e^{-i \omega_{j} t}I_{j}(\tau) \over
\omega \ll f_j|f_j \gg}
  \right] .
\end{equation}
\noindent In (3.1), the source point is $(y,t'=0)$ and the observation point is
$(x,t)$;
for most cases of interest in gravitational systems, $y < x$.  The sum is
over all QNM's
such that the complex frequency $\omega _{j}$ has Re $\omega _{j} > 0$; the
eigenfunctions are $f_{j}(x)$ and $\ll f_{j}\mid f_{j} \gg$ is the
 generalized norm.  The regulating factor $I_{j}$ could be
\begin{equation}
I_{j}(\tau) = e^{-\omega_j \tau},
\end{equation}
\noindent which corresponds to evaluating $G(x,y;t)$ at a complex $t$ with a
small
imaginary part.

We have shown that for discontinuous potentials without a tail, $C1$ holds
if (a) $0 < y < a < x$ and $ t > t_{p} \ge 0$; or
(b) $ 0 < y$ , $x < a$ and $t > 0$.  We further note that with the regulating
factor
$I_{j}(\tau)$, for $(x,t)$ in these two ranges (a) and (b), the sum (3.1) holds
uniformly
and it is valid to differentiate term by term.  This leads to
\begin{equation}
C2: \ \ \ \dot{G}(x,y;t)={\rm Re} \left[ \sum_{j>0}{f_j(x)f_j(y)
 e^{-i \omega_{j} t} I_j(\tau) \over \ll f_j|f_j \gg}\right]  .
\end{equation}
\noindent  $C2$
holds under the same conditions as $C1$, provided a suitable regulating factor
is used.

The importance of $C1$ and $C2$ is that they determine the evolution of initial
data.
For $t>0$, we have
\begin{equation}
\phi (x,t) = \int  dy \left[ G (x,y;t) \dot{ \phi}(y,0)
+ \dot{G} (x,y;t) \phi(y,0) \right].
\end{equation}
The expansion of $G (x,y;t)$ and $\dot{G} (x,y;t)$ in terms of QNM's implies
that
the time development is uniquely determined by the QNM's.  Therefore, for
discontinuous
potentials without tails, if $\phi(y,0)$ and $\dot{ \phi}(y,0)$ have support
only
inside $(0,a)$, $\phi (x<a ,t)$ is completely determined for all times by the
QNM's,
while $\phi (x>a ,t)$ is completely determined by the QNM's for $t >t_p$.

How is this notion of completeness in terms of evolution related to the notion
of completeness in terms of the expansion of the delta function?
We note that from the defining equation and initial condition for
$G$, $\dot{G}(x,y;t=0^+)=\delta(x-y)$;
so if the domain of validity of $C2$ includes $t \rightarrow 0^{+}$, then one
 has in particular
\begin{equation}
C3: \ \ \ \delta(x-y)
={\rm Re} \left[ \sum_{j>0}{f_j(x)f_j(y)  I_j(\tau) \over \ll f_j|f_j \gg}
\right],
\end{equation}
 which may be seen as the more familiar notion of
completeness\cite{convergent}.

It is necessary to distinguish these notions of completeness because,
unlike the familiar case of hermitian systems, completeness in the sense of
evolution ($C1$ and $C2$) is not necessarily equivalent to completeness in
terms of
the expansion of the delta function ($C3$).  This comes about because each of
these holds only in {\it limited} regions of spacetime:  If $y < a <
x$, then $C1$ and $C2$ only hold for $t > t_{p}>0 $;
as $t$ is bounded away from $0 ^ +$, $C3$ may not be valid.
Conversely, suppose we know $C3$ to start with.  In the
familiar case of hermitian systems, such a resolution of the identity
operator $\delta (x-y)$ would allow a decomposition of the initial data into a
sum
of eigenfunctions; attaching phase factors $\exp (-i\omega _{j}t)$ to each of
these
would then give the solution for dynamic evolution.  This procedure
gives an expression for the Green's function $G$ for $t>0$, thus proving $C1$
and $C2.$  This is why completeness in the sense $C3$ is usually regarded as
the key concept for hermitian systems.  However, in the present case,
because $C3$ is valid only inside the ``cavity" $(0,a)$, the superposition of
eigenfunctions would only be valid if $x \in  (0, a)$. Outside this interval,
the initial condition would not be satisfied by a sum of eigenfunctions and
consequently one does not have $C1$ and $C2$ from $C3$ in general.

There is another important case for which $C3$ does not lead to $C1$ and $C2$,
which is useful to sketch here.
Consider the case $V_{o} \equiv m_o^2 \equiv
 \lim_{x\rightarrow \infty} V(x) >  0$.  In this case,
the asymptotic wave number $k$ and the frequency $\omega$ are related by
 $k =\sqrt{\omega^2 - m_{o}^2}$.
The asymptotic behavior would be, for example, $\lim_{x\rightarrow \infty}
\left[ g(\omega ,x) \exp(-ikx) \right ]=1$.  Most of the derivation is easily
adapted, except that the relation between $k$ and $\omega$ leads to an extra
cut $S$ on the real $\omega$ axis.  The integral for $G(x,y;t)$ in terms of
$\tilde{G}$ is above this cut, so in general, there will be this
extra cut contribution when the contour is distorted to the lower half plane.
Thus the QNM sum is {\it not} complete in the sense of $C1$ and $C2$.
Nevertheless, it turns out that this cut
contribution vanishes for $t \rightarrow  0^{+}$, so the QNM representation of
$\delta (x-y)$, i.e., $C3$ remains valid.

\par
\medskip
To see this, define the integral around the cut $S$
\begin{eqnarray}
\nonumber
G_{S}(x,y;t) &&\equiv \int_{S} {d\omega \over 2\pi} \tilde{G}(x,y; \omega )
\exp ( -i \omega t), \\
&&= \int^{{m}_o}_{{-m}_{o}} {d\omega \over 2\pi}
[\tilde{G}(x,y; \omega +i0)-\tilde{G}(x,y; \omega -i0)]\exp (-i\omega t).
\end{eqnarray}
As both $\tilde{G}(x,y; \omega +i0)$ and $\tilde{G}(x,y; \omega -i0)$
satisfy (2.2), their difference  is evidently proportional to the product of
the regular solutions $f(\omega ,x)$ and $f(\omega ,y)$. These
functions diverge exponentially as $x$ or $y$ $\rightarrow \infty$ along  the
real axis, since the asymptotic wave number is now purely imaginary.
However, they are normalizable along the imaginary $x$ or $y$  axis
using the usual box normalization. It can be shown that
$G_{S}$ is expressible as an integral along the imaginary $k$ axis as follows:
\begin{equation}
G_{S}(x,y;t) = \int^{i{m}_{o}}_{0} {\frac{dk}{\pi}} {\hat{f}(\omega,x)
\hat{f}(\omega,y) \cos  \omega t\over \omega },
\end{equation}
\noindent where $\hat{f}(\omega,x)$ is proportional to $f(\omega,x)$
and satisfies the box normalization condition
\par
\begin{equation}
\lim_{L\rightarrow \infty } {1\over iL}\int^{iL}_{0} \hat{f}(\omega,z)^{2} dz =
1.
\end{equation}
\noindent The frequency $\omega $ is now given by
\par
\begin{equation}
\omega  =\mid m^{2}_{o} - k^{2}\mid^{1/2}.
\end{equation}
\noindent It is obvious from (3.7) that
$\dot{G}_{S}(x,y;t=0^+) $ vanishes  and  does  not
contribute to the QNM decomposition of $\delta(x-y)$.

For open systems as studied here, the importance of $C3$ is twofold.
For initial data having support in $(0,a)$, it enables the expansion
of initial data in terms of QNM's.  It is also important in
the construction of a time-independent perturbation
theory in terms of QNM's, an issue we investigated
in the case of optical systems\cite{lly}.

\newpage

\section{Numerical verification}

Consider a potential defined on the full line as follows
\begin{eqnarray}
\nonumber
V(x)&=& V_1 \ \ \ \ {\rm for} \ \ 0 \leq x \leq a,  \\
   &=& 0 \ \ \ \ {\rm otherwise,}
\end{eqnarray}
 with $V_1 > 0.$  This potential has a step discontinuity $( p = 0 )$ and
no tail,
thus satisfying the conditions necessary for our results.  Moreover,
such a repulsive potential has significant leakage, so it
is particularly suited for verifying that our results are not merely valid
to some low order in the amount of leakage.  The QNM's of this system are
determined by
\begin{eqnarray}
\nonumber
f_j(x)&=& {\rm exp}(i \omega x) \ \ \ \ {\rm for} \ x > a ,   \\
\nonumber
     &=& A {\rm exp}(i k x) + B {\rm exp}(- i k x) \ \ \ \ {\rm for} \  a>x>0,
 \\
     &=& D {\rm exp}(- i \omega x) \ \ \ \ {\rm for} \ 0 >x,
\end{eqnarray}
where $ k =\sqrt{\omega^2-V_1} $, and the QNM frequency
$\omega _{j}$ is the $j$-th solution to
\begin{equation}
r(\omega)^2 \exp(2ika) =1,
\end{equation}
with reflection coefficient $r(\omega) \equiv (\omega- k)/(\omega+k) \sim
1/(4 \omega^2)$ at high frequencies. The result (4.3) is a direct
generalization of (2.21) by merely replacing $R(\omega)$ with $r(\omega)^2$,
as waves are now reflected successively by the two discontinuities
at $x=0$ and $ x=a$. The distribution of QNM's
 on the complex $\omega$ plane shown in Fig. 3 corresponds exactly to this
model, and it is easy to show that
\begin{equation}
k_j a \simeq j \pi - 2i[\log(|j| \pi/(V_1^{1/2}a) + \log2]
\end{equation}
for $|j| \gg 1$, consistent with (2.22) with $R \sim
\omega^{-4}$. Im $\omega $ increases with $j$, implying large leakage at high
frequencies.

In the following we shall demonstrate numerically the following
approximate equality for small $\tau$
\begin{equation}
{\rm Re} \sum^{}_{j>0} {f_{j}(x) f_{j}(y)I_j(\tau)\over \ll f_{j} \mid f_{j}\gg
}
\simeq \delta (x-y),
\end{equation}
where $I_j(\tau)=\exp(- \omega_j \tau ) $.
We have labeled the QNM's so that $j > 0$ corresponds to Re $\omega _{j} > 0.$
Since (4.5) is expected to hold only in a distribution sense \cite{convergent},
we integrate $y$ from $y_{1}$ to $y_{2}$
\begin{equation}
{\rm Re} \sum^{}_{j>0} {f_{j}(x) \left[\int^{y_2}_{y_1} dy f_{j}(y)\right]
I_j(\tau)
\over \ll f_{j} \mid f_{j}\gg } \simeq \theta(x-y_1)-\theta(x-y_2).
\end{equation}
Denote the partial sum on the left up to $j = J$ as
$S_{J}(x;y_{1},y_{2};\tau)$.  Fig. 5a shows $|S_{J}|$ versus $J$ for a
case where $x \not \in (y_{1},y_{2})$, and Fig. 5b shows
$\mid S_{J}- 1\mid$ versus $J$ for $x \in (y_{1},y_{2})$.  In each case,
curves are shown for several values of $\tau$.  It is
seen that for sufficiently small $\tau$ (in fact $\tau \ll  \Delta  =
y_{2} - y_{1})$, the equality indeed holds to a good approximation.
These curves also show that convergence in $j$ is more rapid for a slightly
larger $\tau$; this is natural from the regulating factor. This property
implies that in practice, an extremely small $\tau$ may not be optimal.

To study the dependence on $\tau$, Fig. 6a
shows a function $\varphi _{o}(x)$ (e.g., some initial data) on the interval
 $(0,a)$, and
Fig. 6b shows the absolute error in representing this function by the QNM
sum, using $\tau = 10^{-2}$, $10^{-3}$, $10^{-4}$. It is seen that the error
converges to zero as $\tau \rightarrow 0$, and that for most purposes,
using a small finite $\tau$ does not matter in practice. The QNM sum
does not converge if the regulator is removed.

It is useful to examine through this example why the series needs to be
regulated from a broader point of view. Let $r(\omega) \sim \omega^{-q}$
at high frequencies ($q=2$ for the present situation).
By using a similar analysis as that
applied in obtaining (2.22), it is straightforward to show that
Im $k_j a \sim -q \log|j|$. Next consider the middle region $0 < x < a$,
 and a typical term in the product $f_{j}(x)f_{j}(y)$
\begin{equation}
|f_{j}(x)f_{j}(y)| \sim e^{ik_j(x+y)} \sim (j \pi)^{q(|x/a-1/2|+|y/a-1/2|+1)}.
\end{equation}
\noindent  Moreover, it is readily shown that
\begin{equation}
\ll f_{j}\mid f_{j} \gg \ \sim (j \pi)^q.
\end{equation}
\noindent Since the maximum value of $|x/a-1/2|+|y/a-1/2|$ is 1, the worst
behavior of $|f_j(x)f_j(y)|/\ll f_{j}\mid f_{j} \gg$ is $|j|^q $.
It is seen that the sum (4.5) would not converge, even in a distribution sense,
without a regulating
factor if $q \ge  2.$  The culprit is Im $k_{j}a \sim - q \log |j|$,
 from which one
can say unequivocally that the need for regulating the sum is an intrinsic
property of {\it open} KG systems.

 It is worthwhile at this point to reiterate a difference between the wave
equation (WE) and the KG equation.  In each case, if there is a
discontinuity in the {\it p}-th derivative (of the dielectric constant in one
case and of the potential in the other case) then
\begin{eqnarray}
\nonumber
{\rm WE}:\ \ \ && q = p, \\
\nonumber
{\rm KG}:\ \ \ && q = p + 2.
\end{eqnarray}
\noindent Thus in the case of the KG equation, the QNM sum requires regulation
for
all $p \ge  0$, whereas in the case of the WE, regulation is required only for
$p \ge  2$.  This is again natural when seen in the context of (1.10) --- a
step discontinuity in $V$ is equivalent to a second-order discontinuity in
the dielectric constant.  This difference between the two systems, though
of a rather technical nature, should nevertheless be kept in mind when
using one system to draw conclusions about the other.

The need for a singularity is a feature that might seem, at
first sight, to be surprising.  Our results are valid for a
discontinuity however soft (i.e., $p$ however large), but are not valid for a
$ C^{\infty }$
potential.  Yet there should be little difference between a very soft
discontinuity and no discontinuity.  The resolution of this paradox lies in
the regulator --- for large $p$, the divergence is more severe, the regulated
sum converges much more slowly to the correct result as $\tau \rightarrow  0,$
and the use of
a small finite $\tau$ becomes increasingly inaccurate.  This behavior then
connects
smoothly to that for a $C^{\infty }$ potential, for which the sum may not make
sense for
any $\tau$.

\newpage

\section{Discussions and Conclusion}

In this paper we studied the completeness of QNM's for linearized waves
propagating in a curved background described by (1.1).  It is well known that
isolated Schwarzschild black holes do not have a complete set of QNM's,
while other model systems described by (1.1) do have complete
sets of quasinormal modes, e.g., the stellar oscillation model of \cite{price}.
The questions we investigate in this paper are:
(i) What does completeness mean for the QNM's of
an open system described by (1.1)?  (ii) What properties of the potential
would characterize those systems with complete sets of QNM's?
The answers to these questions are spelt out in Secs. II and III.

We find that there are two important ingredients needed for completeness.
First, the potential has to have no ``tail'' in the asymptotic region
in the sense of (2.13). Furthermore, for potentials behaving
asymptotically as $l(l+1)/x^2+\bar{V}(x)$, where $l$ is an integer,
it can be shown that the completeness relation holds provided $\bar{V}(x)$
satisfies (2.13). Details of this generalization is given in other
context\cite{pang}. On the other hand, if $\bar{V}(x)$ decays slower
than an exponential, the QNM's may not form a complete set. It has been
shown that for $\bar{V}(x) \sim 1/x^n$ with $n \ge 2$, the corresponding
Green's function $\tilde{G}(\omega)$ has a branch cut along the negative
imaginary $\omega$ axis, and consequently QNM's are incomplete\cite{prl2}.
The cut is related to the late time tail of gravitational
waves\cite{prl2}, which decays only as an inverse power of time.
This in turn verifies that QNM's do not form a complete set.

Second, the potential has to have discontinuities to provide a
demarcation of a finite interval, analogous to the boundaries of an
optical cavity.  The discontinuities can be in any finite order of the
spatial derivative of the potential.  Discontinuities in the potential
might seem unnatural, however we note that such discontinuities in $V$
commonly exist in many models of gravitating systems, e.g., the Price
potential as a model for the potential of a black hole\cite{chan},
with the discontinuity representing the peak of the potential; and the
stellar oscillation model of \cite{price}, with the discontinuity
representing the stellar surface; as well as other stellar
models\cite{kok1,kok2}.

Needless to say  there are many models of gravitating systems for which
these two conditions for completeness are not satisfied, with perhaps the
most important
example being an isolated Schwarzschild black hole.  For these cases,
we note that our study on completeness is still relevant in
that (i)~it provides
understanding as to {\it why} the quasinormal modes
of these systems should be
incomplete, and (ii)~it shows that their QNM's can
be made complete by a small change in the potential, e.g., by
setting the potential $V(x)$ to zero for $x > X$ for a large $X$ where
$V$ is very small.  On the one hand, one does not expect a
small change to alter the physics much, yet on the other
hand, the set of QNM's now becomes complete and can be a
very useful tool in the analysis of black hole perturbation.
This very interesting point deserves a more careful discussion,
which we now provide.

Consider a smooth potential, labeled schematically as A;
for example this could be the Schwarzschild potential.
Approximate it by a discontinuous potential B, which vanishes
rapidly at spatial infinity; this could, for example,
be a piecewise constant approximation to a  Schwarzschild
potential\cite{nollert}.  Clearly the latter can be chosen such that the
physics of the two systems are nearly the same; schematically PHY(A)
$\simeq $ PHY(B).  Yet it is not difficult to see, e.g., by reference
to the sort of estimates such as (2.22), that the distribution of
QNM's must be fundamentally different, i.e., QNM(A) $\neq $ QNM(B).
Such a qualitative difference has been demonstrated numerically
in the example of a piecewise constant approximation to  Schwarzschild
potential\cite{nollert}. From our point of view, the important difference
lies not only in the distribution of QNM's, but also in that QNM(A)
is not complete, while QNM(B) {\it is} complete.
These differences, both in the asymptotic QNM distribution
and in the completeness, depend on the {\it order} of the discontinuity and
not on the magnitude, and therefore does not go away even when B is
very close to A.  This situation is sometimes regarded as paradoxical
\cite{nollert}. The paradox is
resolved if one realizes that the QNM's are not complete for system A,
and their sum does not correctly describe the physics, i.e., PHY(A)
$\neq $ QNM(A).  Of course, the results in this paper implies PHY(B) =
QNM(B).  Thus, in all, we have
QNM(A) $\neq $ PHY(A) $\simeq$ PHY(B) = QNM(B).

This remark opens up a particularly intriguing possibility.  If we
want to approximate the time development of A by a discrete sum of
QNM's, we should {\it not} use the QNM's of the system
itself. Instead, the QNM's of B, which are complete, can describe the
evolution of A to a good approximation.  Investigations along this
line will be reported elsewhere.

\acknowledgments

We thank H.M. Lai, K.L. Liu, S.Y. Liu, R. Price, and S.S. Tong for
discussions.  This work is supported in part by a grant from the Hong
Kong Research Grant Council (Grant no. 452/95).  The work of WMS is
supported by the US NFS (Grant no. PHY 94-04788) and the CN Yang Fund
while visiting The Chinese University of Hong Kong.

\newpage

\centerline{{\bf FIGURE  CAPTIONS} }

\begin{description}

\item[Fig. 1.] The integrand in (2.10) is defined by the
contour $L = L_1 + L_2$,
where $L_1 = (0,a)$ and $L_2$ is the contour shown in this figure.

\item[Fig. 2.] Contributions from the poles, the semicircle, and the
singularities
of $f$ and $g$ (which in general form a cut on the negative imaginary axis).

\item[Fig. 3.] Distribution of QNM's for the model in Sec. IV with $a=1$ and
$V_1=100$.
\item[Fig. 4.] Contributions from the poles, and the rectangular contour
$\Gamma_n= \Gamma_{n1} \bigcup \Gamma_{n2} \bigcup \Gamma_{n3}$.

\item[Fig. 5.] (a) Plot of $\log |S_J|$ vs $J$ for a case
where $x \not \in (y_1,y_2)$.
(b)  Plot of $\log |S_J-1|$ vs $J$ for a case where $x \in (y_1,y_2)$.
In each case $a=1$, $V_1=100$, $J$ is in thousands, and $S_J$ is the partial
sum in (4.6), evaluated for $\tau =5 \times 10^{-3}$, $5 \times 10^{-4}$,
$2 \times 10^{-4}$ for lines 1, 2, 3 respectively.
The partial sums are in fact fluctuating
functions of $J$, and the lines shown are smooth envelopes representing upper
bounds. These curves show that (i) at fixed $\tau$, the sum over $j$
converges; (ii) the resultant error vanishes as $\tau \rightarrow 0$.

\item[Fig. 6.] (a) A function $\varphi_o(x) = \sin(\pi x/a)$ defined on
$(0,a)$.
(b) The absolute error in representing this function by the QNM sum, using
$\tau=10^{-2}$ (circles), $10^{-3}$ (squares), $10^{-4}$ (triangles).
In this example, $a=1$ and $V_1=100$.

\end{description}


\begin{references}

\bibitem{kip}  K.S. Thorne, Rev. Mod. Phys. {\bf 52}, 299 (1980).
\bibitem{pp}  P.K. Kundu, Proc. R. Soc. London {\bf A431}, 337 (1990);
R.H. Price and J. Pullin, Phys. Rev. {\bf D46}, 2497 (1992).
\bibitem{ppp} It turns out that there is no such suppression,
see R.H. Price, J. Pullin and P. K. Kundu,
Phys. Rev. Lett. {\bf 70}, 1572 (1993).
\bibitem{bhqnm} See, e.g., S. Chandrasekhar, {\it The Mathematical Theory of
Black Holes}, Univ. of Chicago Press (1991).
\bibitem{detweiler}  L. Lindblom and S. Detweiler, Astrophys. J. Suppl.
{\bf 53}, 73 (1983), and references cited therein.
\bibitem{numerwave}  C. V. Vishveshwara, Nature (London) {\bf 227}, 937 (1970);
S. L. Detweiler and E. Szedenits, Astrophys. J. {\bf 231}, 211 (1979);
L. Smarr, in {\it Sources of Gravitational Radiation}, ed. L. Smarr,
Cambridge Univ. Press (1979);
R. F. Stark and T. Piran, Phys. Rev. Lett. {\bf 55}, 891 (1985);
P. Anninos, D. Hobill, E. Seidel, L. Smarr and W.-M. Suen,
Phys. Rev. Lett. {\bf 71}, 2581 (1993); and references therein.
\bibitem{nature} See e.g., A. A. Abramovici et al., Science {\bf 256}, 325
(1992).
\bibitem{prl1}
E.S.C. Ching, P.T. Leung, W.M. Suen and K. Young, Phys. Rev. Lett.
{\bf 74}, 4588 (1995).
\bibitem{price}  R.H. Price and V. Husain, Phys. Rev. Lett.
{\bf 68}, 1973 (1992).
\bibitem{prl2}
E.S.C. Ching, P.T. Leung, W.M. Suen and K. Young, Phys. Rev. Lett.
{\bf 74}, 2414 (1995); E.S.C. Ching, P.T. Leung, W.M. Suen and K. Young,
``Wave Propagation in Gravitational Systems: Late Time Behavior",
to appear in Phys. Rev. D.
\bibitem{lly} P.T. Leung, S.Y. Liu and K. Young, Phys. Rev. {\bf A49},
3057 (1994);
P.T. Leung, S.Y. Liu, S.S. Tong and K. Young,
Phys. Rev. A {\bf 49}, 3068 (1994); P.T. Leung, S.Y. Liu and K. Young,
Phys. Rev. A {\bf 49}, 3982 (1994).
\bibitem{origin} We assume that here $f'$ is nonzero at the origin. Otherwise,
if $f \sim x^{l+1} $ as $x \rightarrow 0$, then we have $\lim_{x \rightarrow 0}
x^{-(l+1)}f(\omega,x) = 1$.
\bibitem{pole}  It is assumed that these are simple poles; generalization to
higher
order poles, or to cuts, is straightforward.  For high order poles, the
time dependence acquires an extra prefactor going like $t^{n}$; see e.g.
J.S. Bell and C.J. Goebel, Phys. Rev. {\bf B138}, 1198 (1965).
\bibitem{zel}  Ya. B. Zeldovich, Zh. Eksp. Teor. Fiz. {\bf 39}, 776 (1960)
[Sov. Phys. - JETP {\bf 12}, 542 (1961)].
\bibitem{lam}  H.M. Lai, C.C. Lam, P.T. Leung and K. Young, J. Opt. Soc.
Am. {\bf B8},
1962 (1991); P.T. Leung and K. Young, Phys. Rev. {\bf A44}, 3152 (1991).
\bibitem{newton}  R.G. Newton, J. Math. Phys. {\bf 1}, 319 (1960).
\bibitem{jordan} Note that Jordan's Lemma states that an
integral $\int d\omega  e^{- i\omega t} f(\omega )$ along a
large semicircle in the lower half plane will vanish for $t > 0,$ provided that
$\left|f(\omega )\right| \rightarrow 0$ at infinity.  The necessary bound
on $f$ arises similarly from
the need to control the part with $\mid \omega _{I}\mid  = O(\log|\omega_R|).$
\bibitem{convergent} To be specific, in this paper we consider
pointwise convergence for $G$ (which is a well defined
quantity) and convergence in the distribution sense for the
representation of $\delta (x-y)$, which comes from $\dot G$.
Going from $G$ to $\dot G$ costs a factor $\omega$, but going
from the pointwise to the distribution
sense effectively gains a factor $\omega ^{-1}$.  This happens
because asymptotically the typical dependence is $e ^{i \omega y}$,
so integration over $y$ gives a factor $\omega ^{-1}$.  Thus
the pointwise validity of (2.12) and the validity of the QNM decomposition of
$\delta$-function in a distribution sense require the same conditions.
\bibitem{0mode} If there are QNM's on the negative imaginary $\omega$ axis,
one has to avoid going over these poles by taking the principal part of the
integral in (2.26). As a consequence, the QNM sum of $G$ in (2.12) and (2.27)
should also include {\it half} of the corresponding sum over those QNM's
with Re $\omega_j =0$.
\bibitem{hardy} G.H. Hardy, {\it Divergent Series}, Oxford, Clarendon
Press (1948).
\bibitem{pang}
P.T. Leung and K.M. Pang, ``Completeness and Time-independent
Perturbation of Morphology-dependent Resonances in Dielectric Spheres",
preprint, Chinese University of Hong Kong (1995).
\bibitem{chan} S. Chandrasekhar and S. Detweiler, Proc. R. Soc. Lond. A.
{\bf 344}, 441 (1975).
\bibitem{kok1} K.D. Kokkotas and B.F. Schutz, Mon. Not. R.
Astron. Soc. {\bf 255}, 119 (1992).
\bibitem{kok2} K.D. Kokkotas,
Mon. Not. R. Astron.  Soc. {\bf 268}, 1015 (1994).
\bibitem{nollert} H.-P. Nollert, ``Quasinormal Frequencies of Step potentials",
preprint, Universit\"{a}t T\"{u}bingen (1994).

\end{references}
\end{document}